\documentclass[
    ,final            
  ,numberedheadings 
  ]
  {aipproc}

\layoutstyle{6x9}

\begin{document}

\title{Mass Varying Neutrinos With More Than One Species Of Neutrinos}


\author{Ole Eggers Bj\ae lde}{
  address={Department of Physics and Astronomy, University of
  Aarhus, Ny Munkegade, Bld. 1520, DK-8000 Aarhus C.} }

\begin{abstract}
In the context of Mass Varying Neutrinos(MaVaNs) we study a model in
which a scalar field is coupled to more than one species of
neutrinos with different masses. In general, adiabatic models of
non-relativistic MaVaNs are heavily constrained by their stability
towards the formation of neutrino nuggets. These constraints also
apply to models with more than one neutrino species, and we find
that using the lightest neutrino, which is still relativistic, as an
explanation for dark energy does not work because of a feedback
mechanism from the heavier neutrinos.
\end{abstract}

\maketitle


\section{Introduction} 
Precise observations of the cosmic microwave background
\citep{Bennett:2003bz,Spergel:2006hy}, the large scale structure of
galaxies \citep{Tegmark:2006az}, and distant type Ia supernovae
\citep{Astier:2005qq} have led to a standard model of cosmology in
which the energy density is dominated by dark energy with negative
pressure, leading to an accelerated expansion of the universe.

A proposal to explain dark energy is the so-called mass varying
neutrino (MaVaN) model~\citep{Hung:2000yg,Gu:2003er,Fardon:2003eh}
in which a light scalar field couples to neutrinos, see also
\citep{Peccei:2004sz,Schrempp:2006mk,Ringwald:2006ks,
Afshordi:2005ym,Brookfield:2005td,Brookfield:2005bz,Takahashi:2006jt,
Spitzer:2006hm,Fardon:2005wc,Kaplinghat:2006jk,Bjaelde:2007ki,
Wetterich:2007kr,Bean:2007nx,Bean:2007ny,Takahashi:2007ru,Mota:2008nj,Bernardini:2008pn,
Ichiki:2008rh,Ichiki:2008st,Valiviita:2008iv, Antusch:2008hj}.

In this paper we discuss the suggestion that the lightest neutrino
which can be relativistic today may be responsible for dark energy.
We find that there is some evidence that the relativistic neutrino
will feel an instability towards the formation of neutrino nuggets.

In the next section we briefly review the formalism needed to study
mass varying neutrinos, in section 3 we discuss MaVaNs with a
relativistic neutrino, and in section 4 we conclude.

\section{Mass Varying Neutrinos} 
In the MaVaN model~\citep{Hung:2000yg,Gu:2003er,Fardon:2003eh} we
introduce a coupling between neutrinos and a light scalar field, and
the coupled fluid then acts as dark energy. In this way, the
neutrino mass $m_\nu$ is generated from the vacuum expectation value
(VEV) of the scalar field. The effective potential is defined by
\begin{equation}
 \centering
 V(\phi) = V_\phi(\phi)+(\rho_\nu-3 P_\nu)
 \label{eq:eff}
\end{equation}
where $V_\phi(\phi)$ is the scalar field potential, $a$ is the scale
factor, $\rho_\nu(m_\nu(\phi),a)$ is the neutrino energy density,
and $P_\nu(m_\nu(\phi),a)$ is the neutrino pressure.

The energy density and pressure of the scalar field are given by the
usual expressions,
\begin{eqnarray}
 \centering
 \rho_\phi(a)=\frac{1}{2a^2}\dot{\phi}^2+V_\phi(\phi)\,\,\,\,\,\,\mbox{and}\,\,\,\,\,\,P_\phi(a)=\frac{1}{2a^2}\dot{\phi}^2-V_\phi(\phi).
\end{eqnarray}

Defining $w=P_{\rm DE}/\rho_{\rm DE}$ to be the equation of state of
the coupled dark energy fluid, where $P_{\rm DE}=P_\nu+P_\phi$
denotes its pressure and $\rho_{\rm DE}=\rho_\nu+\rho_\phi$ its
energy density, the requirement of energy conservation gives,
\begin{equation}
\centering \dot{\rho}_{DE}+3H\rho_{DE}(1+w)=0, \label{eq:Econserv}
\end{equation}
where $H\equiv \frac{\dot{a}}{a}$ and dots to refer to the
derivative with respect to conformal time. Combining with
Eq.~(\ref{eq:Econserv}), one arrives at a modified Klein-Gordon
equation describing the evolution of $\phi$,
\begin{equation}
\ddot{\phi}+2H\dot{\phi}+a^2
V_\phi^{\prime}=-a^2\beta(\rho_\nu-3P_\nu), \label{eq:KG}
\end{equation}
where primes denote derivatives with respect to $\phi$ ($^\prime =
\partial/\partial \phi$) and $\beta=\frac{d {\rm log}m_\nu}{d\phi}$
is the coupling between the scalar field and the neutrinos.

It can be quite instructive to look at the behavior of MaVaN models
in the case of non-relativistic neutrinos $P_\nu\simeq0$, such that
Eq.~(\ref{eq:eff}) takes the form
\begin{equation}
V=\rho_\nu+V_\phi\label{eq:effNR}
\end{equation}

Naturalness suggests we pick a scalar field mass(Curvature scale of
the potential) to be much larger than the expansion rate of the
Universe,
\begin{equation}
\label{eq:mphi}
V^{\prime\prime}=\rho_\nu\left(\beta^{\prime}+\beta^2\right)+V_\phi^{\prime\prime}\equiv
m_\phi^2\gg H^2.
\end{equation}
In this case, the adiabatic solution to the Klein-Gordon
Eq.~(\ref{eq:KG}) applies \citep{Fardon:2003eh}. As a consequence,
the scalar field will sit in the minimum of its effective potential
$V$ at all times
\begin{equation}
 \centering
 V^{\prime}=\rho_\nu^{\prime}+V_\phi^{\prime}=\beta\rho_\nu+V_\phi^{\prime}=0
 \label{eq:phi}
\end{equation}
MaVaNs models can become unstable on sub-Hubble scales
$m_\phi^{-1}<a/k<H^{-1}$ in the non-relativistic regime of the
neutrinos, where the perturbations $\delta\rho_\nu$ evolve
adiabatically.

In Ref.~\citep{Bjaelde:2007ki}(see also
\citep{Hannestad:2005ak,mb,domenico:2002,Koivisto:2005nr,Amendola:2003wa})
it is shown that the equation of motion for the neutrino density
contrast $\frac{\delta\rho_\nu}{\rho_\nu}$ in the regime
$m_\phi^{-1}<a/k<H^{-1}$ can be written as
\begin{equation}
 \ddot{\delta}_\nu
 +H\dot{\delta}_\nu+\left(\frac{\delta
 p_\nu}{\delta\rho_\nu}k^2-\frac{3}{2}H^2\Omega_{\nu}\frac{G_{\rm
 eff}}{G}\right)\delta_\nu=
 \frac{3}{2}H^2\left[\phantom{\frac{|}{|}}\Omega_{\rm CDM}\delta_{\rm
 CDM}+\Omega_{b}\delta_{b}\phantom{\frac{|}{|}}\right]\label{denseom}
\end{equation}
where
\begin{equation}
G_{\rm eff}=G\left(1+\frac{2\beta^2 M^2_{\rm pl}}{1+\frac{a^2}{k^2}
\{V_{\phi}^{\prime\prime}+\rho_\nu\beta^{\prime}\}}\right)\,\,\,\,\,\,\mbox{and}\,\,\,\,\,\,
\Omega_i=\frac{a^2\rho_i}{3H^2M^2_{\rm pl}}\label{Geff}.
\end{equation}
Since neutrinos interact through gravity as well as through the
force mediated by the scalar field, they feel an effective Newton's
constant $G_{\rm eff}$ as defined in Eq.~(\ref{Geff}). The force
depends upon the MaVaN model specific functions $\beta$ and
$V_{\phi}$ and takes values between $G$ and $G(1+2\beta^2 M^2_{\rm
pl})$ on very large and small length scales, respectively.

In certain cases of strong coupling neutrinos suffer an instability
towards clumping in which case they stop behaving as dark energy
\citep{Afshordi:2005ym}. In Ref.~\citep{Bjaelde:2007ki} a criterion
for the stability was developed. $\left(1+\frac{2\beta^2 M^2_{\rm
pl}}{1+\frac{a^2}{k^2}
\{V_{\phi}^{\prime\prime}+\rho_\nu\beta^{\prime}\}}\right)\Omega_\nu\delta_\nu<\Omega_{\rm
CDM}\delta_{\rm CDM}+\Omega_{b}\delta_{b}$. This can be recast in a
more convenient form $\frac{2\beta^2 M^2_{\rm pl}}{1+\frac{a^2}{k^2}
\{V_{\phi}^{\prime\prime}+\rho_\nu\beta^{\prime}\}}\Omega_\nu<\Omega_{\rm
M}$, where we have neglected the effect of baryons compared to cold
dark matter and we have assumed the density contrasts of roughly the
same order.

From the considerations above one can establish a list of criteria
that MaVaN models need to fulfill. This was done in
Ref.~\citep{Bjaelde:2008yd} where it was stressed that for single
field MaVaN models that satisfy adiabaticity, the right amount dark
energy today, correct neutrino mass as well as stability cannot be
simultaneously fulfilled. This has previously been stated by
Refs.~\citep{Afshordi:2005ym} and \citep{Bjaelde:2007ki}.

Hence it has been suggested, in the context of multiple scalar field
models, that neutrinos may be stable towards clustering if our
effective potential has two minima: A false minimum in which our
universe sits and a true minimum. The offset between the two minima
is then interpreted as the dark energy density. The model is
implemented in SUSY to avoid problems with small scalar field
masses. Stability is ensured by letting the lightest relativistic
neutrino be responsible for the dark energy \citep{Fardon:2005wc}.
Below we analyze this suggestion.

\section{MaVaN model with a relativistic neutrino} 

We assume the scalar field couples to all light neutrino species for
naturalness reasons. In the case of three hierarchical neutrino
masses, one would naively assume that, as a result of the coupling,
the two heavier neutrinos would become unstable to clustering,
whereas the lightest would remain stable. In the following we will
argue that this is not possible, since a feedback from the growth of
the heavy neutrino perturbations will cause the relativistic
neutrino perturbation to grow as well. The following equation(see
Ref.~\citep{Brookfield:2005bz}) explains this

\begin{equation}
\centering
 \delta\rho_\nu=\frac{1}{a^4}\int q^2dqd\Omega\epsilon
 f_0(q)\Psi+\delta\phi\beta(\rho_\nu-3P_\nu),
 \label{eq:neutrinop}
\end{equation}
where $\epsilon^2=q^2+m_\nu^2a^2$ and $f_0$ is the unperturbed
Fermi-Dirac distribution with a small perturbation $\Psi$. The
equation explains the growth of the neutrino perturbation and
applies to both the interaction between a relativistic neutrino and
a scalar field as well as that of a non-relativistic neutrino and a
scalar field.

We consider a system consisting of two heavy neutrinos and one
relativistic neutrino each interacting with the same scalar field.
The following list of events will take place. In the beginning, the
relativistic neutrino will not feel the presence of the heavier
ones. However, the heavy neutrinos will feel the coupling which will
drive their perturbations $\delta\rho_\nu(\rm heavy)$ to large
values. In this way, the heavy neutrinos start clumping.

As was demonstrated in Ref.~\citep{Bjaelde:2007ki}, the scalar field
perturbation is effectively proportional to the neutrino
perturbation for the interaction with heavy neutrinos. This means
that since $\delta\rho_\nu(\rm heavy)$ is growing, $\delta\phi$ will
also grow accordingly.

Regarding the relativistic neutrinos, the perturbations
$\delta\rho_\nu(\rm rel)$ will grow as indicated in
Eq.~\ref{eq:neutrinop}. The second term in the equation consists of
three important contributions, firstly $\delta\phi$ which is growing
(this is the same $\delta\phi$ as listed above since we only have
one scalar field). Secondly we have a coupling which we for
simplicity assume to be constant (in reality this will be a growing
quantity for most cases). And finally we have $(\rho_\nu-3P_\nu)$
which is a suppression factor of the order $m/E$ - this factor will
act to delay the growth of $\delta\rho_\nu(\rm rel)$. However, since
$\delta\phi$ will continue its growth, the inevitable conclusion is
that $\delta\rho_\nu(\rm rel)$ will eventually start to grow. Hence,
there exists a type of feedback mechanisms between the heavy and the
relativistic neutrinos. One could of course argue that we are
exactly living in a transition regime when $\delta\rho_\nu(\rm rel)$
has still not turned unstable. However, that would require serious
fine-tuning.

\begin{figure}[!htb]
 \includegraphics[height=.3\textheight]{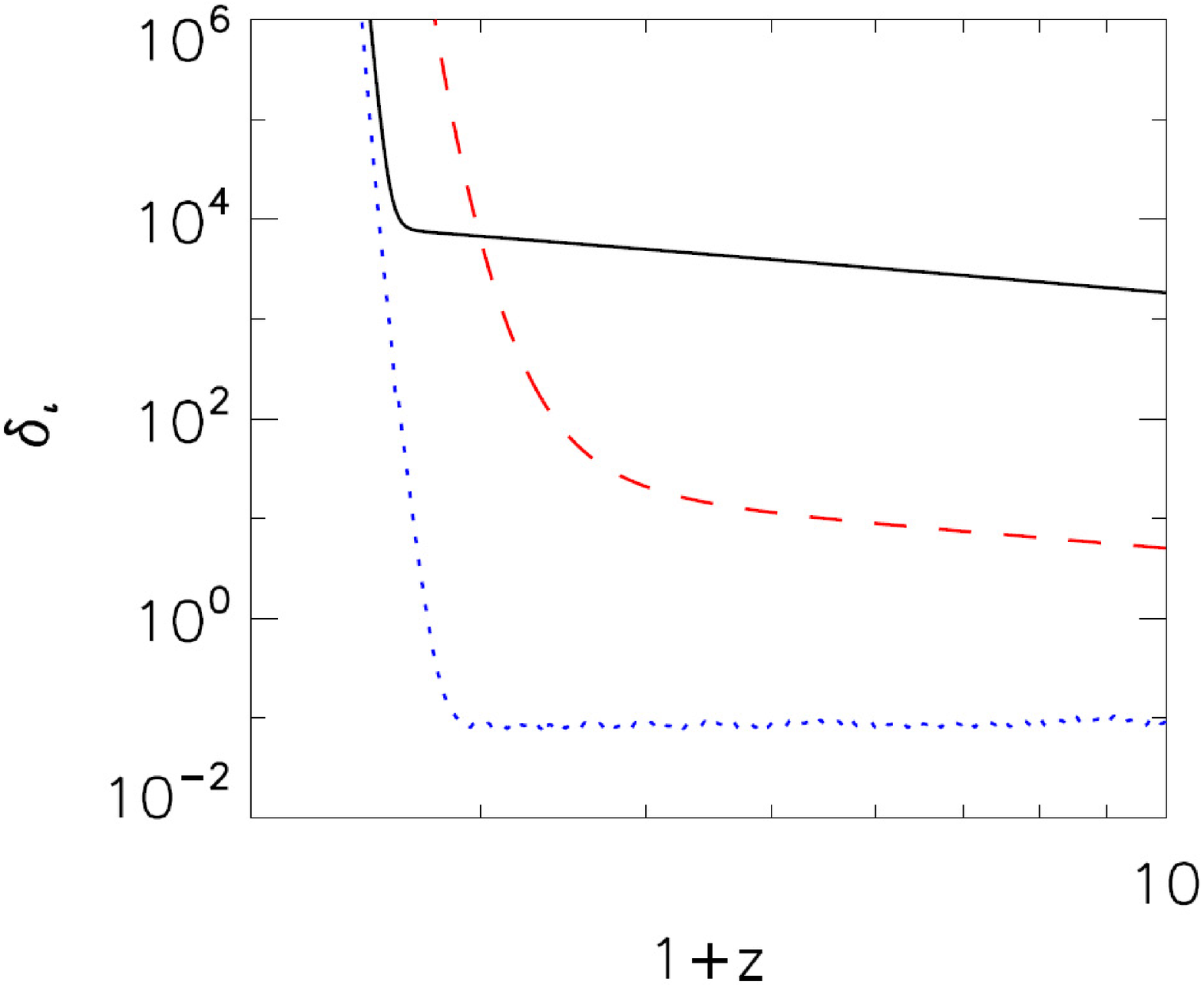}
 \caption{Density contrasts plotted as a function of redshift for a
 system consisting of one light and two heavy MaVaN neutrinos each
 interacting with the same scalar field. The scale is $k=0.1\,{\rm
 Mpc}^{-1}$ and we choose the current neutrino masses
 $m_\nu(\rm rel)=10^{-7}$\,eV and $m_\nu(\rm heavy)=0.3$\,eV (Note that the choice
 of current neutrino masses does not affect the result
 qualitatively). The solid line is cdm-density contrast, the dotted
 line is the light neutrino density contrast, and the dashed line is
 the heavy neutrino density contrast. The heavy neutrinos grow
 essentially as cdm until the coupling becomes large enough for the
 instabilities to set in. The light neutrino is still relativistic,
 and its density contrast oscillates as acoustic waves. However, due
 to a feedback mechanism, the relativistic neutrino density contrast
 tracks that of the non-relativistic neutrino around the time the
 growth of the heavy neutrino perturbations become quasi-linear. I.e.
 both neutrino species will clump. Note that the cdm perturbations
 also blow up at late times. This is an effect of the system of
 differential equations breaking down as all parameters go to
 infinity.} \label{fig:dens}
\end{figure}

A graphical illustration of the example above is given in
Fig.~\ref{fig:dens}. This is done in the framework of a model with a
Coleman-Weinberg type scalar field potential similar to the one
presented in Refs.\citep{Fardon:2003eh} and \citep{Bjaelde:2007ki}.
\begin{equation}
 \centering
 V_\phi=V_0\,\log{\left(1+k^2\phi^2\right)}
\end{equation}
In order to avoid possible pathological behaviour, we choose a mass
term slightly different than the one in Ref.~\citep{Bjaelde:2007ki},
namely one that does not become infinite when the VEV of the scalar
field goes to zero. However, it still behaves as $1/\phi$ for small
$\phi$.
\begin{equation}
 \centering
 m_\nu=-\frac{1}{2}\lambda\phi+\sqrt{\frac{1}{4}\lambda^2\phi^2+m_d^2},
\end{equation}

which can be derived from solving the mass matrix
$(\lambda\phi,m_d;m_d,0)$.

What happens is shown in Fig.~\ref{fig:dens}, where we can see that
for the higher redshifts the density contrast of heavier neutrinos
behaves moderately as predicted by GR. The cdm term in
eq.~\ref{denseom} sources the slow growth of these heavy neutrinos.
As their masses increase more and more the coupling term slowly
takes over and becomes the dominant term in eq.~\ref{denseom}.
Eventually this leads to the unstable growth of their density
contrast. What happens next is that once the growth of the heavy
neutrinos enter the quasi-linear regime, immediately the effect can
be seen on the growth of the relativistic neutrino density contrast.
This starts blowing up, and a short while later the system of
equations we are solving effectively breaks down, which can be seen
by the unnatural strong growth of the cdm density contrast.

This gives us a hint that as a result of a feedback mechanism, the
fast growing behavior of the heavy neutrino density contrast causes
the relativistic neutrinos to start clumping as well.

Hence the neutrino scalar-field fluid will start acting as a cold
dark matter component (clustering neutrinos) and hence cannot be
attributed to dark energy.

Now, one can of course argue that the rise in the density contrast
of the relativistic neutrino species only happens once the heavier
ones have turned non-linear. In this regime the linear code does not
apply. However, from a close-up look at the data, we emphasize that
the rise of the density contrast happens in the quasi-linear regime
of the heavy neutrino perturbations, where the code does still
apply.

The reason that the relativistic neutrino is able to clump is that
it will acquire an effective mass, thus it cannot be regarded as a
relativistic particle. Unfortunately, we cannot use conventional
bounds to constrain this effect for the following reason: Once the
evolution of the non-relativistic neutrinos becomes non-linear, the
whole system of equations we are solving, starting with the modified
KG equation breaks down. This has the effect that all current bounds
are no longer valid, as these are established in the linear regime.

\section{Conclusion} 
Single scalar field models can be used to explain late-time
acceleration in the MaVaN scenario. However, in general using these
potentials leads to instabilities towards neutrino bound states
unless certain criteria are relaxed.

Accordingly it has been suggested to include an extra scalar field
in the treatment. This has some very nice features and is easily
capable of obtaining late-time acceleration as well as
$\Omega_{\mathrm{DE}}=0.7$ today. However, one drawback is the need
for the lightest neutrino to be relativistic today. As was explained
above, the feedback mechanism will eventually cause the relativistic
neutrino to start clustering and hence the coupled fluid will cease
to act as dark energy.

\begin{theacknowledgments}
OEB would like to thank Steen Hannestad for great supervision, and
Lily Schrempp, Carsten van de Bruck, Anthony Brookfield, David Mota,
and Domenico Tocchini-Valentini for collaboration. OEB acknowledge
the use of the publicly available CMBFAST code \citep{CMBFAST}.

This is a contribution to the conference proceedings from the 4th
International Workshop on the Dark Side of the Universe 2008 in
Cairo and will appear in AIP Conference Proceedings 1115 \cite{aip}.
\end{theacknowledgments}

\end{document}